\documentclass[reprint, twocolumn, aps, amsmath, amssymb, prd]{revtex4-2}

\usepackage{comment}
\usepackage{verbatim}
\usepackage{bm}
\usepackage[dvipsnames]{xcolor}
\usepackage{fullpage}
\usepackage{graphicx}
\usepackage{booktabs}
\usepackage{array}
\usepackage{CJKutf8}
\usepackage[colorlinks = true,
            linkcolor = blue,
            urlcolor  = blue,
            citecolor = blue,
            anchorcolor = blue]{hyperref}
            
\usepackage{orcidlink}
            
\newcommand{\be}{\begin{equation}}
\newcommand{\ee}{\end{equation}}
\newcommand{\bea}{\begin{eqnarray}}
\newcommand{\eea}{\end{eqnarray}}
\def\ba#1\ea{\begin{align}#1\end{align}}

\def\({\left(}
\def\){\right)}

\def\[{\left[}
\def\]{\right]}

\newcommand{\refeq}[1]{Eq.~(\ref{eq:#1})}
\newcommand{\refeqs}[2]{Eqs.~(\ref{eq:#1})--(\ref{eq:#2})}
\newcommand{\reffig}[1]{Fig.~\ref{fig:#1}}
\newcommand{\refsec}[1]{Sec.~\ref{sec:#1}}
\newcommand{\refapp}[1]{App.~\ref{app:#1}}
\newcommand{\reftab}[1]{Tab.~\ref{tab:#1}}
\newcommand{\vs}{\nonumber\\}

\def\vx{\bm{x}}
\def\vq{\bm{q}}

\def\vk{{\bm{k}}}

\def\bfu{\bm{u}}

\def\d{{\rm d}}
\def\<{\left <}
\def\>{\right >}
\def\dD{\delta^D}

\def\dn#1{\delta^{(#1)}}
\def\tn#1{\theta^{(#1)}}
\def\un#1{\bfu^{(#1)}}

\def\nn{\nonumber}

\definecolor{RedWine}{rgb}{0.743,0,0}
\definecolor{GrassGreen}{rgb}{0.125,0.75,0.125}
\definecolor{RoyalBlue}{rgb}{0.25,0.41,0.88}

\def\mnras{Mon.\ Not.\ R.\ Astron.\ Soc.}

\def\aap{Astron.\ Astrophys.}
\def\apj{Astrophys.\ J.}
\def\aj{Astron.\ J.}
\def\apjs{Astrophys.\ J. Supp.}

\def\prd{Phys.\ Rev.\ D}

\def\physrep{Phys. Rep.}

\def\jcap{{JCAP\ }}
\def\pasj{PASJ}

\begin{document}

\title{Perturbation Theory Remixed II: Improved Modeling of Nonlinear Bispectrum}

\author{
Zhenyuan Wang (\begin{CJK*}{UTF8}{gbsn}王震远\end{CJK*})*\orcidlink{0000-0002-2970-3661},$^{1}$ 
Donghui Jeong\orcidlink{0000-0002-8434-979X},$^{1,2}$
Atsushi Taruya\orcidlink{0000-0002-4016-1955},$^{3,4}$\\
Takahiro Nishimichi\orcidlink{0000-0002-9664-0760},$^{5,3,4}$ and
Ken Osato\orcidlink{0000-0002-7934-2569}$^{6,7,4}$
}

\affiliation{%
$^{1}$ 
Department of Astronomy and Astrophysics and Institute for Gravitation and the Cosmos, 
The Pennsylvania State University, University Park, PA 16802, USA
\\
$^{2}$ School of Physics, Korea Institute for Advanced Study, Seoul, South Korea
\\
$^{3}$ Center for Gravitational Physics and Quantum Information, Yukawa Institute for Theoretical Physics, Kyoto University, Kyoto 606-8502, Japan
\\
$^{4}$ Kavli Institute for the Physics and Mathematics of the Universe, Todai Institutes for Advanced Study, the University of Tokyo, Kashiwa, Chiba 277-8583, Japan 
\\
$^{5}$ Department of Astrophysics and Atmospheric Sciences, Faculty of Science, Kyoto Sangyo University, Motoyama, Kamigamo, Kita-ku, Kyoto 603-8555, Japan
\\
$^{6}$ Center for Frontier Science, Chiba University, Chiba 263-8522, Japan
\\
$^{7}$ Department of Physics, Graduate School of Science, Chiba University, Chiba 263-8522, Japan
}

\email{zzw173@psu.edu}
\date{\today}

\begin{abstract}
{
We present the application of the $n$-th order Eulerian Perturbation Theory ($n$EPT) for modeling the matter bispectrum in real space as an advancement over the Standard Perturbation Theory (SPT). The $n$EPT method, detailed in Wang et al. (2023) \cite{Wang2023nEPT}, sums up the density perturbations up to the $n$-th order before computing summary statistics such as bispectrum. Taking advantage of grid-based calculation of SPT (GridSPT), we make a realization-based comparison of the analytical nonlinear bispectrum predictions from $n$EPT and SPT against a suite of $N$-body simulations. Using a spherical-bispectrum visualization scheme, we show that $n$EPT bispectrum matches better than SPT bispectrum over a wide range of scales in general $w$CDM cosmologies. Like the power spectrum case, we find that $n$EPT bispectrum modeling accuracy is controlled by $\sigma_8(z) \equiv \sigma_8 D(z)$, where $D(z)$ is the linear growth factor at a redshift $z$. Notably, the 6EPT doubles the bispectrum model's validity range compared to the one-loop SPT for $\sigma_8(z) < 0.5$, corresponding to redshifts $z\ge1$ for the best-fitting Planck-2018 cosmology. For $n\ge5$, however, $n$EPT bispectrum depends sensitively on the cut-off scale or the grid resolution. The percent-level modeling accuracy achieved for the spherical bispectrum (where we average over all triangular configurations) becomes much degraded when fixing configurations. Thus, we show that the validity range of the field-level cosmological inferences must be different from that derived from averaged summary statistics such as $n$-point correlation functions.
}\end{abstract}


\maketitle


\section{Introduction}
The large-scale structure of the universe is one of the major observables for extracting cosmological information. The leading-order statistics of large-scale structure, the two-point correlation function, has been the main summary statistics for analyzing, for example, the anisotropies of temperature and polarization of the cosmic microwave background (CMB) radiation measured from WMAP \cite{Bennett2013WMAP} and Planck \cite{Aghanim2020Planck} satellites, and the distribution of galaxies measured from the galaxy redshift surveys such as Sloan Digital Sky Survey 
 and Dark Energy Spectroscopic Instrument survey \citep{Flaugher2014}. 

For further cosmological studies, for example, to determine the initial condition of the Universe, to study the nature of dark matter and dark energy, and to test gravity on large scales, many large-scale structure surveys with wider sky coverages and deeper flux limits are currently undergoing or being planned. These include HETDEX \cite{gebhardt/etal:2021}, Euclid \citep{Laureijs2011}
LSST of Vera Rubin Telescope \citep{Ivezic2019}, Subaru PFS \citep{Takada2014PFS}, HAWAS (High-Altitude Wide-Area Survey) of the Roman Space Telescope \citep{Spergel2015}, and SPHEREx \citep{Dore2014}.

While the large-scale structure traced by the CMB radiation is measured to be nearly Gaussian \citep{Planck2018fNL}, that traced by galaxies is genuinely non-Gaussian due to the nonlinearities in the late-time gravitational evolution, the galaxy formation and evolution, and the redshift-space distortion. Therefore, analyzing only the two-point correlation function misses opportunities to extract the non-Gaussian cosmological information \citep{Cheng2020, Hahn2022SimBIG}, and fully exploiting the galaxy clustering dataset demands analyses beyond the two-point correlation function \citep{beyond2pt}.

The bispectrum, the Fourier equivalent of the three-point correlation function, is the lowest-order statistics sensitive to non-Gaussianities in galaxy clustering \citep{Sefusatti2006}. Incorporating the galaxy bispectrum into the analysis disentangles parameter degeneracies in the power-spectrum-only analysis, for example, between linear galaxy bias and $\sigma_8$, and significantly tightens the constraints on cosmological parameters \citep{Agarwal2021, Ivanov2023}, including, but not limited to, the neutrino mass \citep{Hahn2020, Hahn2021, Yankelevich2023} and primordial non-Gaussianity \citep{Dizgah2021}. 

Standard Perturbation Theory (SPT) provides the foundational tool for modeling the gravitational nonlinearities in the cosmic density field. Treating the matter field on large scales as pressureless perfect fluid, that is, neglecting velocity dispersion and higher-order velocity moments in BBGKY hierarchy, PT solves the Vlasov--Poisson equation perturbatively in the order of linear density contrast. Then, the SPT solution is used to predict the ensemble mean of the summary statistics, such as the power spectrum \citep{Jeong2006} and bispectrum \citep{Fry1984, Scoccimarro1997, Scoccimarro1998} on the quasi-linear scales.
The most advanced analytical SPT bispectrum calculations are at the next-to-leading (NLO, or one-loop) order because higher-order loop corrections are tied to complex, high-dimensional integrals, making calculations computationally challenging \citep{Slepian2020, Philcox2022bk}. 
Going beyond the NLO, at the same time, SPT is also known for its poor convergence when adding successively higher-order corrections to the power spectrum \citep{Blas2014, Konstandin2019}. With these limitations and the back-reaction of small-scale nonlinearities to large scales \citep{Pueblas2009} that is not captured in SPT, the application of SPT is limited to modeling nonlinearities in high redshifts.

To improve the accuracy of SPT modeling and remedy pathological convergence behavior for higher-order SPT, beginning with \cite{Crocce2006RPT}, a plethora of renormalized PT schemes have been proposed over the past two decades. These include resummed Lagrangian perturbation theory \citep{Matsubara2008}, renormalized perturbation theory (RPT) \citep{Crocce2006RPT, Crocce2006memory} including ${\rm MPT_{BREEZE}}$ \citep{Crocce2012, Lazanu2016} and Regularized Perturbation Theory (RegPT) \citep{Taruya2012}, to name a few. Alternatively, the Effective Field Theory of Large-scale Structure (EFTofLSS) \citep{Baumann2012, Carrasco2012, Hertzberg2014} method is proposed to absorb the small-scale back-reactions into a few counter terms; each counter term in the model for the nonlinear power spectrum and bispectrum introduces a free parameter, which must be determined along with the cosmological parameters. Although the details of the performance deviate method by method, these efforts successfully extend the range of model validity beyond SPT. We often quantify the validity range by $k_{\rm max}$, which is the maximum wavenumber below which a theory accurately models the nonlinearities.

Both renormalized PT and EFTofLSS methods have also been used to model the nonlinearities in the matter bispectrum. The success of resummed methods in increasing the model's validity range for nonlinear power spectrum, however, (see also \citep{Osato2019, Osato2023}) has not been extended much in modeling bispectrum \citep{Lazanu2016, Alkhanishvili2022}, and only modest improvement has been obtained. While the validity range of the EFTofLSS for bispectrum modeling is, in general, larger than SPT and renormalized PT \citep{Carrasco2014,Lazanu2016,Angulo2015,Baldauf2015b}, the precise value of $k_{\rm max}$ depends on many factors, including the configurations used for comparison, the total simulation volume which determines the statistical uncertainties, the range of wavenumber to determine the EFT free parameters and systematics by mass resolutions in $N$-body simulations \citep{Alkhanishvili2022}. For example, the EFT parameters significantly deviate from its low-$k$ limit when fitted from the bispectrum in $N$-body simulation with $k > 0.12\, h$/Mpc at $z = 1$ \citep{Alkhanishvili2022}.

In this paper, we present the nonlinear bispectrum model using the novel $n$EPT ($n$-th order Eulerian Perturbation Theory) \citep{Wang2023nEPT} method, which is proven to extend the $k_{\rm max}$ well beyond the SPT for the nonlinear matter power spectrum. While using the same $n$-th order density contrast computed using the SPT recursion relation, $n$EPT differs from the standard PT practice in the way it computes the summary statistics. Namely, $n$EPT first adds the nonlinear contributions up to a fixed order $n$ and then calculates the summary statistics using the total density contrast. In particular, we compute the $n$-th order density fields from the grid-based calculation of SPT (GridSPT) \citep{Taruya2018Grid} with a given linear density field. GridSPT allows a realization-by-realization comparion of the $n$EPT result with the $N$-body simulations starting from the exactly same linear density field \citep{Wang2023nEPT}.
Ref.~\citep{Wang2023nEPT} also shows that $n$EPT power spectrum enjoys well-behaved convergence for successively increasing $n$, so long as the nonlinearities are not too strong (for the redshifts $z \ge 1$). One indication is that $k_{\rm max}(n)$ of power spectrum steadily increases as a function of the PT order $n$. Furthermore, 5EPT is proven to outperform the two-loop SPT power spectrum at $z \ge 1$ and reach a similar level of accuracy as RegPT+ and EFT, but without employing {\it any} free parameters. 

A similar approach using the Lagrangian Perturbation Theory has been studied in \cite{Schmidt2021,Stadler/etal:2023} and well demonstrated the powerfulness of the method. These Lagrangian studies, however, typically focus on much larger scales by employing a somewhat conservative cutoff scale $\Lambda\lesssim 0.2 \, h/{\rm Mpc}$. Our Eulerian method complements the Lagrangian studies on finding the extension of the modeling to smaller scales by including the higher-order contributions.

The success of $n$EPT in modeling the power spectrum may open a new avenue for the field-level modeling of the nonlinear density field using $n$EPT scheme. To explore this possibility further, we have to check the accuracy of the $n$EPT modeling beyond the power spectrum, which is the averaged amplitude modulus of the density field in Fourier space. In this paper, we check the accuracy of the $n$EPT modeling for the nonlinear bispectrum, again in a realization-by-realization manner, by taking full advantage of the GridSPT \citep{Taruya2018Grid} method. This comparison is more stringent than the ensemble-mean-based comparison usually adopted to test the accuracy range of the summary statistics of SPT, renormalized PT, and EFTofLSS.

Being the three-point correlation function of density contrast, the bispectrum depends on three wavevectors that form a triangle in the Fourier space. Even after exploiting the underlying statistical isotropy, we are left with three parameters $k_1$, $k_2$, and $k_3$, making visualization for bispectra comparison hard. To facilitate the comparison between the $n$EPT and the $N$-body results, we use the spherical bispectrum \citep{Tomlinson2023} to reorganize the bispectrum information according to the spherical wavenumber of the Fourier triangles, $k_{\rm sph} \equiv \sqrt{(k_1^2+k_2^2+k_3^2)/3}$, with which we can project the bispectrum amplitude onto one-dimension.

At the end, we confirm that $n$EPT's accurate modeling of the nonlinear density field holds up for the bispectrum as well as the power spectrum \citep{Wang2023nEPT}. The main finding of the analysis is as follows. First, the $n$EPT has better convergence behavior than SPT: the accuracy of $n$EPT bispectrum model is excellent on large scales (lower wavenumber) and deviates from the $N$-body results at increasingly smaller scales as we include higher-order contributions to the density field. This behavior is in contrast to the sign-alternating residual curves common in the SPT model (for example, see the three right panels of Fig. 1 in \cite{Wang2023nEPT} or \reffig{WMAP} in this paper). Furthermore, the improvement of $k_{\rm sph, max}$, the maximum wavenumber defined with 2\% residual of spherical bispectrum, of $n$EPT compared to SPT is even larger than that in the nonlinear power spectrum we have reported in \cite{Wang2023nEPT}. We also find a tight anti-correlation between $k_{\rm sph, max}$ and $\sigma_8(z)$, which characterizes the amplitude of the linear density field and provides a common ground to compare the result at different redshift with various cosmologies. We also confirm that the statement holds up for general $w$CDM cosmologies around the best-fitting Planck2018 cosmology.

The rest of this paper is organized as follows. We introduce the bispectrum modeling with $n$EPT and SPT in \refsec{method}, the bispectrum visualization in \refsec{visualization}, and the setup of $N$-body simulations in \refsec{nbody}.
Then, we compare the $n$EPT bispectrum against the $N$-body results in \refsec{result}. We conclude and discuss future work in \refsec{conclusion}. In \refapp{bijkoptimal}, we present a fast method of measuring the density bispectrum from the GridSPT output.

Throughout the paper, we use the following convention of Fourier transformation,
\bea
&&\tilde f(\vk) = \int \d^3 x f(\vx) e^{-i\vk\cdot \vx},
\\
&&f(\vx) = \int \frac{\d^3 k}{(2\pi)^3} \tilde f(\vk) e^{i\vk\cdot \vx},
\eea
but remove the tilde of the Fourier mode in the rest of the paper, as they are barely a different projection of the same Hilbert-space function: $f(\vk)=\left<\vk|f\right>$, $f(\vx)=\left<\vx|f\right>$. For the compactness of the equations, we use the following convention for the sum of multiple vectors,
\bea
\vk_{1\cdots n} = \sum_{i=1}^n \vk_i.
\eea

\section{Perturbation Theory model of nonlinear bispectrum}
\label{sec:method} 
\subsection{Overview: Standard Perturbation Theory (SPT)}
\label{sec:SPT}
The Standard Perturbation Theory (SPT) treats the large-scale evolution of dark matter and baryon perturbations as that of a pressureless perfect fluid, which follows the set of fluid equations:
\ba
&\dot{\delta} + \nabla\cdot\left[(1+\delta){\bm v}\right] = 0 ,
\\
&\dot{\bm v} + ({\bm v}\cdot\nabla){\bm v} + \frac{\dot a}{a}{\bm v} = - \nabla\phi ,
\ea
along with the Poisson equation:
\ba
\nabla^2 \phi = 4\pi G \bar{\rho}_\mathrm{m}\, a^2 \delta.
\ea 
Here, $\delta$ is the density contrast $\delta=\rho_{\rm m}/\bar{\rho}_{\rm m}-1$ and ${\bm v}=\dot{\vx}$ is the peculiar velocity. Note that the dot represents the conformal-time derivative ($d\tau=dt/a$, with the scale factor $a(t)$ and cosmic time $t$), and $\vx$ is the comoving coordinate in the FLRW spacetime, and $\nabla$ is the comoving-coordinate derivative. Finally, $\phi$ is the peculiar gravitational potential. 

Furthermore, SPT \citep{bernardeau2002large} assumes irrotational velocity at all orders and expands the density contrast $\delta$ and the reduced velocity-divergence $\theta\equiv -\(\nabla\cdot{\bm v}\)/(aHf)$ as 
\bea
\delta(\tau,{\bm x})= \sum_n [D(\tau)]^n\delta^{(n)}({\bm x}),
\label{eq:def_delta}
\\
\theta(\tau,{\bm x})= \sum_n [D(\tau)]^n\theta^{(n)}({\bm x}).
\label{eq:def_theta}
\eea
In Fourier space, the spatial parts $\dn n(\vk)$ and $\tn n(\vk)$ can be written as
\bea
\dn n(\vk) = \int_{\vq_1}\cdots\int_{\vq_n}(2\pi)^3 \dD{\(\vk-\vq_{1\cdots n}\)} 
\nn\\
\times F_n(\vq_1,\cdots,\vq_n) \prod_i \delta_L (\vq_i),
\label{eq:deltan}
\\
\tn n(\vk) = \int_{\vq_1}\cdots\int_{\vq_n}(2\pi)^3 \dD{\(\vk-\vq_{1\cdots n}\)} 
\nn\\
\times G_n(\vq_1,\cdots,\vq_n) \prod_i \delta_L(\vq_i),
\label{eq:thetan}
\eea
with the linear density field $\delta_L$ and the density and velocity kernels $F_n$ and $G_n$, respectively. Here, $\delta^D$ is the Dirac delta operator.
For general spacetime, the kernels depend on time and satisfy the following  differential equations \cite{Jeong:thesis}:
\ba
&\frac{1}{f(\dot{a}/a)}\dot{F}_n + nF_n-G_n = \sum_{m=1}^{n-1}\frac{\vk_{1\cdots n}\cdot\vk_{1\cdots m}}{k_{1\cdots m}^2}G_m F_{n-m}\,,
\label{eq:FGeq1}
\\
&\frac{1}{f(\dot{a}/a)}\dot{G}_n 
+ \(\frac32\frac{\Omega_{\rm m}}{f^2}+n-1\)G_n
-
\frac32 \frac{\Omega_{\rm m}}{f^2}F_n 
\vs
=& \sum_{m=1}^{n-1}
\frac{k_{1\cdots n}^2(\vk_{1\cdots m}\cdot\vk_{m+1\cdots n})}{2k_{1\cdots m}^2k_{m+1\cdots n}^2}G_m G_{n-m}\,,
\label{eq:FGeq2}
\ea
but the standard practice is to compute the kernels for the Einstein de-Sitter (EdS; flat, matter-dominated) Universe where $\Omega_{\rm m}=f=1$ where the kernels are time-independent. That is, we fix the kernels to their EdS form, and the cosmology dependence appears as the time dependence coming through the linear growth factor $D(\tau)$. 
This prescription is proven to be accurate at a sub-percent level for $\Lambda$CDM cosmologies \citep{takahashi2008third, Fasiello2022} (see Ref~\citep{Schmidt2021} for such comparison in Lagrangian perturbation theory).

\subsection{Grid-based Calculation of Standard Perturbation Theory (GridSPT)}
\label{sec:GridSPT}
In EdS universe, the recursion relation for $F_n$ and $G_n$ are algebraic equations, \refeqs{FGeq1}{FGeq2} without the time-derivative term. Even with that, however, implementing the higher-order perturbative solutions in Fourier space requires multi-dimensional convolution of \refeqs{deltan}{thetan} \citep{Roth2011}. GridSPT \citep{Taruya2018Grid} bypasses the difficulties by using the configuration-space recursion relation:
\begin{align}
\label{eq:recursion}
\begin{pmatrix}
\dn n(\vx) \\
\tn n(\vx)
\end{pmatrix}
= \frac{1}{(2n+3)(n-1)} \begin{pmatrix}
2n+1 & 1 \\
3 & n
\end{pmatrix} 
\nonumber\\
\times  \sum_{m=1}^{n-1} \begin{pmatrix}
\nabla \cdot \[\dn m(\vx)\un{n-m}(\vx)\] \\
\nabla^2 \[\un{n-m}(\vx) \cdot \un{n-m}(\vx)\]
\end{pmatrix}\,.
\end{align}
Here, ${\bm u}\equiv -{\bm v}/aHf$ is the reduced peculiar velocity, which is related to $\theta$ by $\un n(\vk) = - i\vk\tn n(\vk)/k^2$. Note that we set the DC ($\vk=0$) mode to be zero, ${\bm u}(\vk=0) = \tn n(\vk=0) = 0$ to avoid any effect from the long-wavelength modes.

Equipped with the Fast Fourier Transformation (FFT) algorithm, GridSPT generates SPT nonlinear density contrast and velocity field from a given linear density contrast quite efficiently for both real-space \citep{Taruya2018Grid, Taruya2021Grid} and redshift-space \citep{Taruya2022RSD}. The computational complexity reduces from $\mathcal{O} (N^2)$ \citep{Roth2011} to $\mathcal{O} (N \log N)$, where $N$ stands for the total number of grids.

\subsection{$n$-th order Eulerian Perturbation Theory ($n$EPT)}
Unlike the traditional PT-based methods that compute the statistical quantities up to a given total PT order, $n$EPT first computes the sum of the density contrast up to $n$-th order to estimate the statistical quantities from the total. That is, $n$EPT constructs the density contrast in the Eulerian space from the first to $n$-th order, which we denote as $\delta_{n\rm EPT}$, then calculate the summary statistics of $\delta_{n\rm EPT}$.

In the strict SPT framework, the two methods must converge in the limit of $n\to\infty$. The series of assumptions on which SPT is built, however, must also break down well before that limit. In the literature, the limitation of SPT is explored in relation to traditional PT methods. For example, see \citep{Baldauf2015a, Baldauf2016} for the EFT correction of two-loop order power spectrum and \citep{Angulo2015, Baldauf2015b, Steele2021, Aghanim2020Planck} for the one-loop bispectrum to improve the accuracy of SPT at low redshifts. The limitation of $n$EPT way of modeling nonlinear density field has never been done before. Therefore, the precise question we address here is how far we can push the quasi-linear scales ($k_{\rm max}$) by employing the $n$EPT method.

We contrast the difference between SPT and $n$EPT in computing power spectrum and bispectrum as follows. The $n$EPT power spectrum and bispectrum
\ba
\label{eq:nEPTpk}
&P_{n{\rm EPT}}(k;\tau) 
=
\sum_{i,j \le n} P_{ij}(k) D^{(i+j)}(\tau)
\\
&B_{n{\rm EPT}}(k_1, k_2, k_3;\tau) 
\vs
&=
\sum_{i,j,k \le n} B_{ijk}(k_1, k_2, k_3) D^{(i+j+k)}(\tau),
\label{eq:nEPTbk}
\ea
which contains nonlinear contributions different from the $n$-loop power spectrum and bispectrum in SPT:
\ba
\label{eq:SPTpk}
&P_{\rm SPT}^{n\text{-}{\rm loop}}(k;\tau)
= 
\sum_{i+j \le n+2} P_{ij}(k) D^{(n+2)}(\tau)
\\
\label{eq:SPTbk}
&B_{\rm SPT}^{n\text{-}{\rm loop}}(k_1,k_2,k_3;\tau)
\vs
= &
\sum_{i+j+k \le n+4} B_{ijk}(k_1, k_2, k_3) D^{(n+4)}(\tau)\,.
\nn\\
\ea
Here, we have defined
\ba
&\< \delta^{(i)}({\bm k_1})\delta^{(j)}({\bm k_2})\> \equiv (2\pi)^3P_{ij}(k_1)\delta^D(\vk_{12})\,,
\\
&\< \delta^{(i)}({\bm k_1})\delta^{(j)}({\bm k_2})\delta^{(k)}({\bm k_3})\>
\vs
& \equiv 
(2\pi)^3B_{ijk}(k_1,k_2,k_3)\delta^D(\vk_{123})\,.
\ea
One crucial difference between the traditional SPT method and $n$EPT is the inclusion of the {\it odd} components in summary statistics: $P_{ij}$ and $B_{ijk}$ when $i+j$ or $i+j+k$ are odd numbers. These contributions are neglected in the traditional SPT methods because their ensemble mean vanishes for Gaussian linear density fields. When comparing the power spectrum and bispectrum for a single realization of GridSPT and $N$-body simulations, however, these odd components make a non-zero contribution to the statistics, and including the odd components indeed enhances the accuracy of the $n$EPT prediction. In addition, these odd components also contribute to the statistical uncertainties of measured summary statistics. In fact, fixed-pair simulation suites \citep{Angulo2016, Pontzen2016, villaescusa-navarro2018fixpair} cancel out the odd components and significantly decrease the statistical uncertainties. That is the reason behind the good match between the fixed-pair simulation suites simulation suites and traditional SPT predictions.

\section{Bispectrum Visualization}
\label{sec:visualization}
Correlating three density contrasts in Fourier space, the matter bispectrum can, in principle, be a function of three wavevectors ${\vk}_1$, ${\vk_2}$, and ${\vk}_3$ with nine degrees of freedom (dof) in total. However, statistical homogeneity reduces the dof to six by demanding the triangular condition ${\vk}_{123}=\bm{0}$. Note that any three vectors subject to the triangular condition must be on a plane, but statistical isotropy demands that the matter bispectrum must be independent from the direction of the plane ($-2$ dof) and the rotation on the plane ($-1$ dof). The matter bispectrum, therefore, can be fully characterized by the rest of the three dof. Visualizing the three-dimensional dependences is challenging as it would require a four-dimensional display to fully encapsulate all information (e.g. see \citep{Lazanu2016}).
 
 One common way of specifying the three dof is to use the norm of three wave vectors, represented as $(k_1, k_2, k_3)$. Earlier works \cite{Scoccimarro1997,Scoccimarro1998,Verde/etal:1998} often show only a part of the triangular shapes in one-dimensional plot. When showing the full dependences, for example, Ref.~\citep{Scoccimarro2000} has condensed the three wavenumbers into a one-dimensional array of flattened indices, with each index corresponding to a specific triangle configuration. \reffig{bkwmapflatz1var} shows an example. While spreading the bispectrum along the flattened index allows a two-dimensional display of the bispectrum in its entirety, it often fails to isolate key features in the bispectrum, such as its scale and configuration dependencies, and, thereby, obstructs the physical interpretation.

To enhance the clarity of bispectrum comparisons between theory and $N$-body results, we adopt the spherical-bispectrum visualization technique \citep{Tomlinson2023} detailed in \refsec{bksph}. This approach highlights the scale dependence of the bispectrum by collapsing the configuration dependence. The spherical-bispectrum visualization suits well the main goal of this paper of assessing the $k_{\rm max}$ of the $n$EPT bispectrum. 

Once we determine $k_{\rm max}$, we shall reflect our findings using the flattened-index visualization to check the configuration dependence of the $k_{\rm max}$. For constructing the flattened index consistent with the spherical-bispectrum visualization, however, we have invented a new permutation strategy, ordering the flattened index so that each flattened-index chunk corresponds to the same $k_{\rm sph}$, different from the usual practice aligning each chunk with the same $k_1$. We present the details of the new flattening index in \refsec{bkflatksph}.

\subsection{Spherical-bispectrum visualization}
\label{sec:bksph}

The Cartesian coordinate consisting of the triplet of wavenumbers $(k_1, k_2, k_3)$ can be written in the corresponding spherical coordinate $(k_{\rm sph}, \theta_{\rm sph}, \phi_{\rm sph})$ as
\bea
\label{eq:ksph}
k_{\rm sph} &=& \sqrt{(k_1^2 + k_2^2 + k_3^2)/3}
\\
\theta_{\rm sph} &=& \arctan\({k_3 / \sqrt{k_1^2 + k_2^2}}\)
\\
\phi_{\rm sph} &=& \arctan{\(k_1/k_2\)}.
\eea

The spherical bispectrum refers to binning the bispectrum according to the radius in the spherical coordinate $k_{\rm sph}$ of each Fourier triangle $(k_1, k_2, k_3)$. This binning scheme allows to show the matter bispectrum as a one-dimensional function of $k_{\rm sph}$, facilitating the comparison between various perturbation theory outcomes to the $N$-body results to determine the maximum spherical wavenumber $k_{\rm sphp, max}$ of each theory model.

In practice, we first measure both the bispectrum $B(k_1, k_2, k_3)$ and the number of triangles $N_B(k_1, k_2, k_3)$ that belong to each spherical-bispectrum bin: $|\sqrt{k_1^2+k_2^2+k_3^2}-k_{\rm sph}|<\delta k_{\rm sph}/2$. Then, the spherical bispectrum in each $k_{\rm sph}$ bin is obtained by the weighted average:
\bea
B(k_{\rm sph}) = \frac{\sum B(k_1, k_2, k_3) N_B(k_1, k_2, k_3)}{\sum N_B(k_1, k_2, k_3)}\,.
\eea
Here, we sum over the triangles with spherical wavenumber falling into the bin of $k_{\rm sph}$. Throughout this paper, we set the bin size to be the fundamental wavenumber $k_F = 2\pi/L$, where $L$ is the side length of $N$-body box.

Let us consider the range of $k_{\rm sph}$ for a fixed largest wavenumber ($k_1\ge k_2\ge k_3$). From the triangular condition, we can find the minimum $k_{\rm sph}$ as
\ba
&k_1^2 =  \,|\vk_2+\vk_3|^2 \le k_2^2+k_3^2+2k_2k_3 
\vs &\quad \le \,2\(k_2^2+k_3^2\) = 2\(3k_{\rm sph}^2-k_1^2\)
\vs &\to\,\frac{k_1}{\sqrt2}\le k_{\rm sph}\,.
\ea
where the second inequality comes from the Cauchy--Schwarz inequality: $a^2+b^2\ge 2ab$. The equality holds when $k_2=k_3=k_1/2$ and $\vk_2\parallel\vk_3$, which is the case for the folded triangles. The maximum $k_{\rm sph}$ comes about for the equilateral triangle ($k_1=k_2=k_3$) where $k_{\rm sph}=k_1$. Therefore, for a fixed $k_1$, the spherical wavenumber $k_{\rm sph}$ is limited by
\be
\frac{k_1}{\sqrt{2}}\le k_{\rm sph} \le k_1\,.
\ee
In other words, to complete the spherical bispectrum up to some maximum spherical wavenumber $k_{\rm sph, \mathrm{max}}$, we have to measure the bispectrum up to $k_1=\sqrt{2}\,k_{\rm sph, \mathrm{max}}$.

When fixing triangular configurations, the spherical bispectrum only averages over the corresponding corners in the $(k_1,k_2,k_3)$ Cartesian coordinate. In such cases, the required wavenumber range to complete the spherical bispectrum also varies with the triangular configurations. Interested readers can find the requirements in Tab. 1 of \cite{Tomlinson2023}.

\subsection{Flattened Index with fixed $k_{\rm sph}$}
\label{sec:bkflatksph}

\begin{figure}[h!t]
    \centering
    \includegraphics[width=0.495\textwidth]{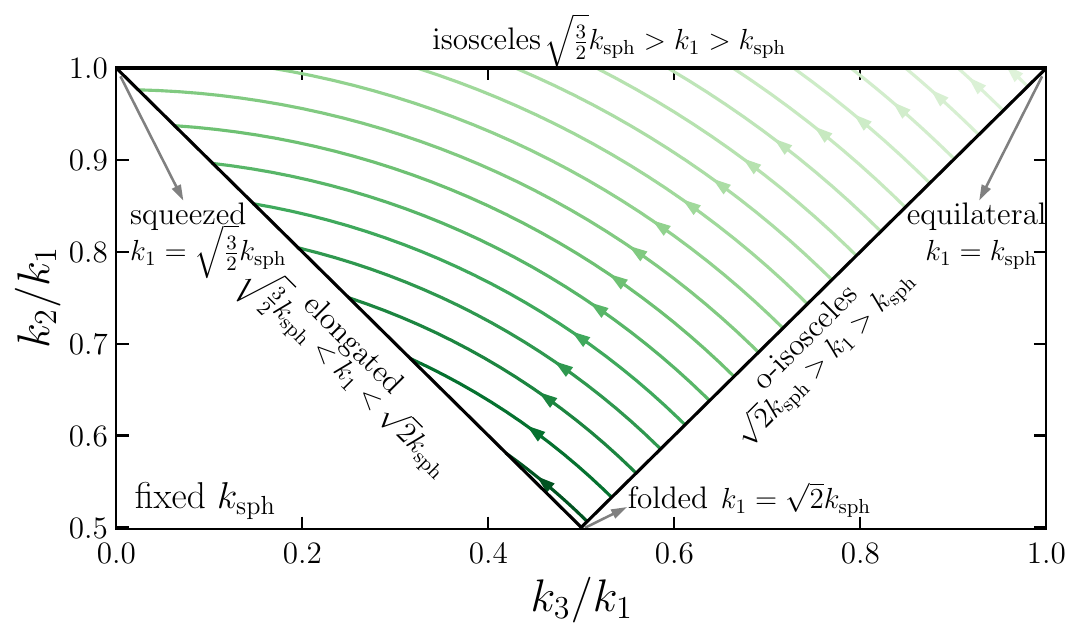}
    \caption{Illustration of constructing the flattened index for sampled Fourier triplets ($k_1, k_2, k_3$), where $k_1 \ge k_2 \ge k_3$. Sampling of triangles begins along the lighter green lines and progresses to the darker green lines. Within each line, triangles are sampled in the direction indicated by the arrows.
    We first fix $k_{\text{sph}}$ and then sample triangles with all possible $(k_1, k_2, k_3)$ that correspond to the fixed $k_{\text{sph}}$. Each green line corresponds to a unique $k_1$ value between $k_1$ and $\sqrt2 k_1$, as indicated in the plot. We highlight the range of $k_{1}$ for special triangle configurations, including folded, elongated, squeezed, isosceles, o-isoseles, and equilateral triangles.} 
    \label{fig:flatindex}
\end{figure}

The usual approach to constructing the flattened index is alphabetically ordering $(k_1,k_2,k_3)$ triplet ($k_1\ge k_2\ge k_3$). That is, for a fixed $k_1$, varying $(k_2,k_3)$ doublet from $(k_2,k_3)=(1/2,1/2)k_1$ (folded triangle) to $(k_2,k_3)=(k_1,k_2)$ (equilateral triangle).

When further decomposing the spherical bispectrum into the triangular configurations, however, the flattened index constructed in this manner is not compatible with the spherical bispectrum, since triangles within the same $k_{1}$ chunk may fall into different bins of spherical wavenumber $k_{\rm sph}$. For example, $k_{\rm sph} = k_1/\sqrt{2}$ for folded triangles and $k_{\rm sph} = k_1$ for equilateral triangles.

To remedy the situation, we construct the new flattened index scheme by first fixing $k_{\rm sph}$ then sampling $(k_1, k_2, k_3)$ alphabetically for those triplets falling in the same $k_{\rm sph}$ bin. As a result, the points on the fixed-$k_1$ plane belong to different chunks depending on their $k_{\rm sph}$ values, as illustrated in \reffig{flatindex}. For a fixed $k_{\rm sph}$, the quadratic sum of $k_{2}$ and $k_{3}$ follows
\bea
\(\frac{k_{2}}{k_{1}}\)^2 + \(\frac{k_{3}}{k_{1}}\)^2 = 3\(\frac{k_{\rm sph}}{k_{1}}\)^2 - 1,
\eea 
which are arcs with radius varies with $k_{\rm sph}/k_{1}$. The new sampling method naturally extends the spherical bispectrum to highlight the triangular configuration dependencies. \reffig{WMAPflat} shows one such example.

\subsection{Triangle Configurations}
In \refsec{bkflat}, we test the $n$EPT modeling of bispectrum for different triangle configurations, including equilateral, isosceles, o-isosceles, squeezed, elongated, folded, and general triangles, whose definitions are visulized in \reffig{flatindex}. However, some triangle configurations following the strict definition could result in very limited number of modes and stochastic bispectrum result.

To alleviate this problem, we use the following relaxed definition of triangle shapes in \refsec{bkflat}, following \citep{Tomlinson2023}, 
\begin{itemize}
    \item Equilateral: $k_1 = k_2 = k_3$
    \item Isosceles-like: $0.95< k_2/k_1 < 1$ or $k_3/k_2 > 0.95$
    \item Squeezed-like: $k_1/k_3 > 3, k_2/k_1 > 0.95 $
    \item Elongated-like: $k_1 > 0.95(k_2+k_3)$
    \item General: all the other triangles
\end{itemize}
Here the o-isosceles triangles are categorized into isosceles-like ones, and elongated-like triangles also include folded triangles for the clarity of the plot. We also list the fraction of the triangle numbers of each configuration for $k_{\rm sph} < 0.25 \, h$/Mpc.
\begin{itemize}
    \item Equilateral: $1.01\%$
    \item Isosceles-like: $22.1\%$
    \item Squeezed-like: $3.04\%$
    \item Elongated-like: $8.28\%$
    \item General: $66.3\%$
\end{itemize}
For simplicity, we omit the "-like" suffix in the configuration names in the rest of this paper.

\section{Setup: $N$-body simulations, GridSPT \& Bispectrum Estimator}
\label{sec:nbody}
\renewcommand{\arraystretch}{1.3}
\begin{table*}[h!t]
\caption{The cosmological parameters used for the baseline (WMAP run) and 20 Dark-Quest run simulations.}
\small
\centering
\begin{tabular}{m{3cm} m{1.7cm} m{1.7cm} m{1.7cm} m{1.7cm} m{1.7cm} m{1.7cm} m{1.7cm}}
\toprule
\; \; \;Name          & \; $n_s$  & \; $h$    &\; $\Omega_b$&\; $\Omega_m$&\; $\Omega_w$& \; $w$    &\; $\sigma_8$ \\
\midrule
\; \; WMAP          &  0.96  & 0.701  & 0.046    & 0.279    & 0.721    & -1     & 0.8159   \\
\hline
Dark Quest 1  & 0.9447 & 0.6971 & 0.0437   & 0.3115   & 0.6885   & -0.842 & 0.8851  \\
Dark Quest 2  & 0.9293 & 0.6315 & 0.0534   & 0.3635   & 0.6365   & -1.074 & 0.6500  \\
Dark Quest 3  & 0.9592 & 0.8937 & 0.0269   & 0.1883   & 0.8117   & -1.114 & 1.0480  \\
Dark Quest 4  & 0.9544 & 0.6064 & 0.0585   & 0.4046   & 0.5954   & -0.918 & 0.9662  \\
Dark Quest 5  & 0.9717 & 0.5783 & 0.0647   & 0.3936   & 0.6064   & -1.126 & 0.6990  \\
Dark Quest 6  & 0.9930 & 0.8051 & 0.0336   & 0.2129   & 0.7871   & -0.982 & 0.7210  \\
Dark Quest 7  & 0.9168 & 0.6067 & 0.0594   & 0.3772   & 0.6228   & -0.946 & 0.8926  \\
Dark Quest 8  & 0.9216 & 0.8620 & 0.0296   & 0.1938   & 0.8062   & -1.022 & 0.8204  \\
Dark Quest 9  & 1.0036 & 0.5553 & 0.0716   & 0.4347   & 0.5653   & -0.870 & 0.6105  \\
Dark Quest 10 & 0.9862 & 0.6557 & 0.0517   & 0.3443   & 0.6557   & -1.014 & 0.6401  \\
Dark Quest 11 & 0.9602 & 0.7839 & 0.0364   & 0.2513   & 0.7487   & -0.966 & 0.7179  \\
Dark Quest 12 & 0.9766 & 0.6364 & 0.0552   & 0.3334   & 0.6666   & -1.146 & 1.0569  \\
Dark Quest 13 & 0.9332 & 0.6834 & 0.0482   & 0.2951   & 0.7049   & -1.182 & 1.0502  \\
Dark Quest 14 & 0.9688 & 0.5676 & 0.0704   & 0.4237   & 0.5763   & -0.822 & 0.9021  \\
Dark Quest 15 & 0.9496 & 0.7800 & 0.0375   & 0.2348   & 0.7652   & -0.802 & 0.8035  \\
Dark Quest 16 & 0.9795 & 0.6910 & 0.0479   & 0.3170   & 0.6830   & -0.894 & 1.1437  \\
Dark Quest 17 & 1.0016 & 0.7022 & 0.0465   & 0.2677   & 0.7323   & -0.934 & 0.8600  \\
Dark Quest 18 & 0.9370 & 0.7998 & 0.0361   & 0.2212   & 0.7788   & -1.094 & 0.7855  \\
Dark Quest 19 & 1.0094 & 0.5768 & 0.0696   & 0.4456   & 0.5544   & -1.058 & 0.7272  \\
Dark Quest 20 & 0.9978 & 0.7246 & 0.0444   & 0.2759   & 0.7241   & -1.170 & 0.8757  \\
\midrule
\end{tabular}
\label{tab:cosmology}
\end{table*}

We test the performance of $n$EPT in modeling the nonlinear bispectrum by comparing the result of SPT and $n$EPT against two groups of $N$-body simulation that we call WMAP run and Dark-Quest run in what follows. Since we use the GridSPT (\refsec{GridSPT}) realization to forward-model the nonlinear density field from the {\it exact} initial linear density field used to run the $N$-body simulations (also see \citep{Taruya2018Grid, Schmidt2021, Baldauf2021, Wang2023nEPT}), we can make such a comparison even with a single realization. This method must be contrasted with the comparison between perturbation theory and ensemble average of $N$-body simulation results (for example, see \citep{Jeong2006, Osato2019, Tomlinson2023, Alkhanishvili2022}). Here, the variation that we study among the WMAP run and Dark-Quest run is the underlying cosmological model. Also, our implementation for SPT or $n$EPT here does not contain any free parameters so that the obtained validity range is not impacted by the uncertainties of fitted parameter due to, for example, limited simulation volume, which typically happens for theories like EFTofLSS that involves free parameters \citep{Alkhanishvili2022}.

The WMAP run uses an $N$-body simulation of $1024^3$ particles and of box size $L = 1\;{\rm Gpc}/h$, with the cosmological parameters consistent with the WMAP-5 results \citep{komatsu2009five}. We use the dark-matter distribution data at redshifts $z = 0$, $0.5$, $1$, $2$, $3$, and $5$ for our base-line comparison. 

The Dark-Quest run consists of 20 $N$-body simulations in Dark Quest Project \citep{Nishimichi2019DQ} with different stochastic initial conditions. The 20 simulations are for the 20 test cosmologies arranged uniformly over the six-dimensional hyperrectangle, covering roughly up to a $\sim$10$\sigma$ range of the 2015 Planck data \citep{Planck2015}. Each simulation in Dark Quest uses $2048^3$ particles, and the box size is the same as the WMAP run $L = 1\;{\rm Gpc}/h$. 

The $N$-body simulation data we use here is the same as the ones to test the performance of the $n$EPT power spectrum in \cite{Wang2023nEPT}. The cosmological parameters we use to perform GridSPT calculation and $N$-body simulations can be found in \reftab{cosmology}.

For both WMAP and Dark-Quest runs, we compute the Fourier-space density contrasts by using the Julia implementation of FFTW of the dark matter particle density assigned on the regular $1024^3$ grid by the CIC (cloud-in-cell) scheme. We also correct for the aliasing effect by employing the interlacing method \citep{Sefusatti2016}. The measurement error is much less than 1\% up to the Nyquist frequency of $k=3.2 \, h/\mathrm{Mpc}$. 

For the GridSPT calculation, we compute the nonlinear density contrast up to sixth-order in PT, which is sufficient for computing two-loop SPT and 6EPT bispectrum using \refeq{SPTbk} and \refeq{nEPTbk}, respectively. To avoid the aliasing effect when using the recursion relation in \refeq{recursion}, we use the generalized Orszag rule \citep{Schmidt2021, Taruya2022RSD}: we compute the GridSPT on $N_{\rm grid}^3$ regular grid, and the linear density field for the wavevector $|\vk|>[2/(n+1)]k_{\rm Nyq}$ is set to be zero (called zero padding) for $n$EPT. Here, $k_{\rm Nyq}=\pi/(L/N_{\rm grid})$ is the Nyquist wavenumber. The requirement $k_{\rm cut}^{\rm UV}<[2/(n+1)]k_{\rm Nyq}$ for a desired $k_{\rm cut}^{\rm UV}$ sets the minimum $N_{\rm grid}$ that we use for the GridSPT calculation. For the baseline computation, we use the fiducial UV cutoff $k_{\rm cut, fid}^{\rm UV} = 256k_{F} =  1.61\, h$/Mpc, which sets $N_{\rm grid}={\rm 1792}$. We also use two other UV cutoffs $k_{\rm cut}^{\rm UV} \in (200, 340)k_F = (1.26, 2.14)\,h$/Mpc to study the sensitivity of the modeled bispectrum to the UV cutoffs (the grid resolution). This sets the $N_{\rm grid} \in (1400, 2380)$.

Now we have the Fourier-space density grid from the WAMP run, the Dark-Quest run, and corresponding GridSPT calculations. We finally measure the bispectrum using the paralleled polyspectra estimator with slab-decomposed paralleled FFT algorithm \citep{Tomlinson2019}, which computes the bispectra $B(n_1k_F,n_2k_F,n_3k_F)$ and number of triangular configurations $N_B(n_1k_F,n_2k_F,n_3k_F)$ for each triplet $(n_1,n_2,n_3)$ of non-negative integers and the fundamental wavenumber $k_F=2\pi/L$. The detailed steps of measurement of SPT and $n$EPT bispectrum from GridSPT density perturbations can be found in \refapp{bijkoptimal}. Then, we organize the full bispectrum dataset into spherical bispectrum according to \ref{sec:bksph}, with the the fundamental wavenumber $k_F$ as the bin size $k_{\rm sph}$. Since both GridSPT and $N$-body analyses follow the precisely same steps, the binning corrections \citep{Eggemeier2019, Oddo2020, Tomlinson2023} do not impact the comparison here.  We also subtract the Poissonian shot noise from the measured bispectrum in $N$-body simulations, but the effect of shot noise is negligible. We compare the measured bispectrum from $n$EPT and $N$-body simulations at 21 redshifts from $z = 0$ to $1.48$.

\section{Results: Bispectrum Comparison}
\label{sec:result}
\begin{figure*}[h!t]
    \centering
    \includegraphics[width=2\columnwidth]{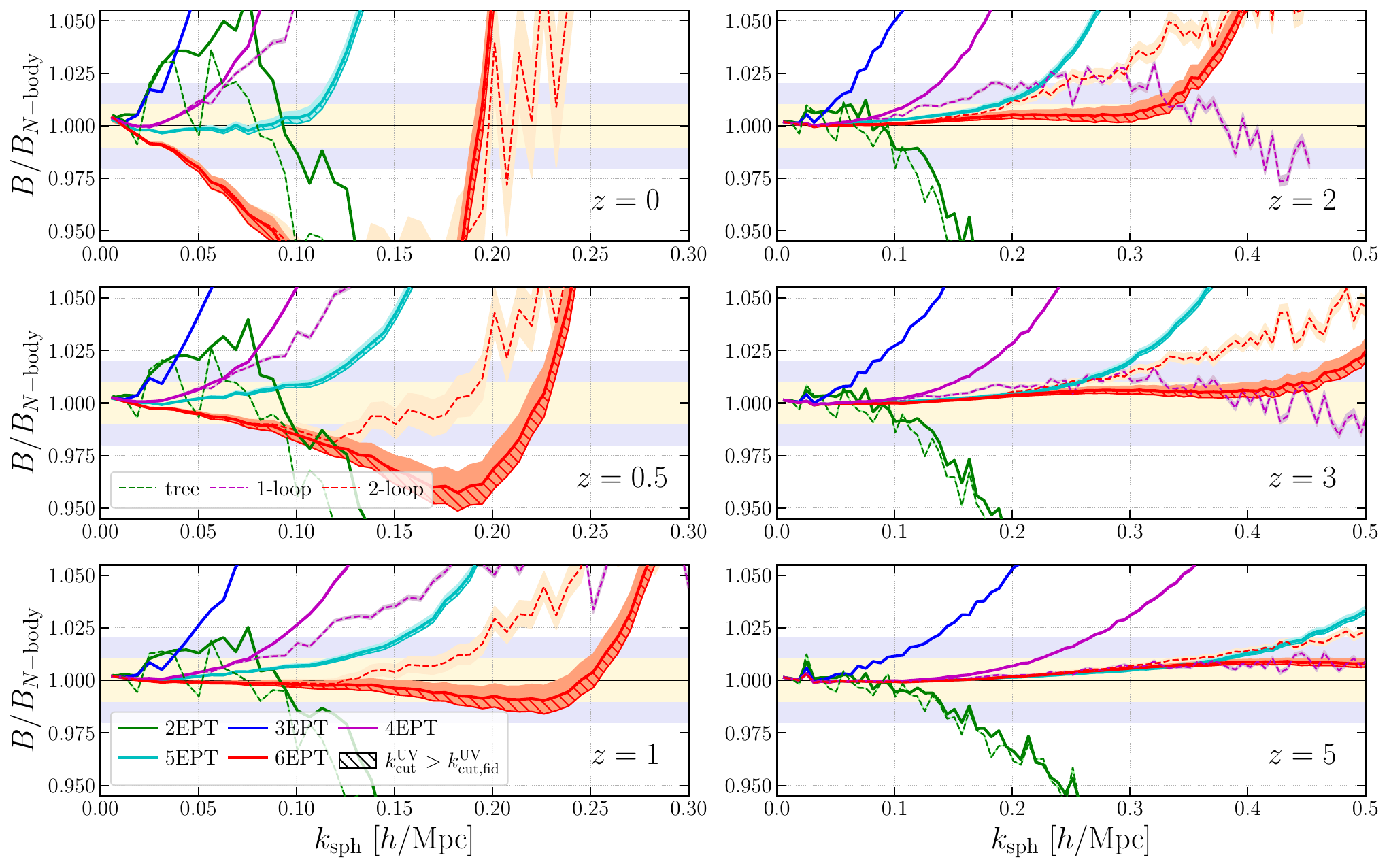}
    \caption{The ratio of spherical bispectrum modeled by $n$EPT to $N$-body results for the baseline WMAP5 cosmology at redshifts $z = 0,\, 0.5,\, 1,\, 2,\, 3,\,$ and $5$. The thick solid lines are the result from $n$EPT calculations: 2EPT (green), 3EPT (blue), 4EPT (magenta), 5EPT (cyan), and 6EPT (red). The think dashed lines are the result from SPT calculations: tree (green), one-loop (magenta), and two-loop (red). Both $n$EPT and SPT results are measured from the density field forward modeled by GridSPT using the initial linear density field of $N$-body simulation. The shaded regions show the variation with UV cutoff between $(1.26, 2.14)$ $h$/Mpc. The central lines are from the results with fiducial UV cutoff $k_{\rm UV, fid} = 1.61$ $h$/Mpc. The hatched area indicates the variation with higher UV cutoff than the fiducial value. The yellow and lavender bands indicate the $\pm$1\% and $\pm$2\% regions.}
    \label{fig:WMAP}
\end{figure*}
In this section, we compare the matter bispectrum computation of SPT and $n$EPT using GridSPT against the $N$-body results. Here, the `SPT bispectrum' refers to the usual tree-level, one-loop, and two-loop calculation as described in \refeq{SPTbk}, and the `$n$EPT bispectrum' refers to the model detailed in \refeq{nEPTbk}. In \refsec{bksphWMAP}, as a baseline study, we compare the spherical bispectrum from $n$EPT and SPT for the WMAP run adopting the $\Lambda$CDM cosmology. We then extend this comparison to the Dark-Quest run adopting the general $w$CDM cosmologies, as discussed in \refsec{bksphDQ}. Finally, in \refsec{bkflat}, we scrutinize the precision of $n$EPT bispectrum modeling by spreading the bispectrum averaged in the $k_{\rm sph}$ bin to the flattened index showing the comparison at each $(k_1,k_2,k_3)$ triplet.

\subsection{Spherical Bispectrum in $\Lambda$CDM Cosmology: WMAP run}
\label{sec:bksphWMAP}
\reffig{WMAP} shows the ratio of SPT (dashed lines) and $n$EPT (solid lines) spherical bispectrum to the baseline $N$-body results at six redshifts: $z = 0$, $0.5$, $1$, $2$, $3$, $5$. 
The yellow and lavender bands in each panel highlight the one and two-percentage error ranges. Note that the ranges of spherical wavenumber $k_{\rm sph}$ are different: the left three panels ($z=0$, $0.5$, $1$) shows for $k_{\rm sph} \le 0.3\,h$/Mpc, and the right three panels ($z=2$, $3$, $4$) shows for $k_{\rm sph} \le 0.5\,h$/Mpc

\reffig{WMAP} clearly demonstrates several advantages of $n$EPT over SPT in bispectrum modeling.

First and foremost importantly, $n$EPT can model the bispectrum accurately for a much wider range than SPT at high redshifts, $z \ge 1$. For example, 6EPT (red solid line) bispectrum follows the $N$-body result to a sub-percentage accuracy up to $k_{\rm sph}=0.25\,h$/Mpc at $z = 1$, while two-loop SPT bispectrum (red dashed line) crosses the one-percent line at $k_{\rm sph}=0.19\,h$/Mpc. Notably, the current most advanced analytical SPT bispectrum calculation, the one-loop (purple dashed line) SPT calculation, achieves a percent accuracy only up to $k_{\rm sph} = 0.11\,h$/Mpc. That is, compared to the one-loop SPT, using 6EPT expands $k_{\rm max}$ by more than a factor of two, with which we expect immense gain in the statistical information: This corresponds to a factor of $(25/11)^3\simeq 11.7$ increase in the total number of triangular configurations, each of which contains the factor $11.7$ more $N_B$. The enhancement provided by 6EPT over the two-loop SPT bispectrum is more substantial than the improvement that 5EPT yields in the power spectrum compared to the two-loop SPT \citep{Wang2023nEPT}.

Second, successively higher order $n$EPT spherical bispectrum shows better convergence than that of SPT at $z\ge2$. Specifically, before reaching the $k_{\rm max}$, $n$EPT spherical bispectrum's agreement with $N$-body results improve consistently as the order $n$ increases. In contrast, the SPT spherical bispectrum exhibits poor convergence from one-loop to two-loop. For example, the one-loop SPT spherical bispectrum (dashed-purple lines) maintains accuracy within $\pm 2\%$ up to $k = 0.5\; h$/Mpc at $z = 3$ (the middle left panel of \reffig{WMAP}). However, while the addition of two-loop correction (dashed red lines) slightly improves the accuracy at smaller $k$, it reduces the accuracy for $k \ge 0.3\; h$/Mpc. With such a behavior in convergence of SPT, it is reasonable to doubt that the good match between one-loop SPT and $N$-body result beyond $k \simeq 0.3\; h$/Mpc at $z=3$ is rather a result of serendipity than an outcome of robust modeling. We can observe similar patterns for $z = 2$ and $z = 5$ cases.

Finally, in \reffig{WMAP}, the residuals of the $n$EPT bispectra (for $n \ge 4$) are much smoother than the residuals of the SPT bispectra. As explained earlier, we attribute the smoothness to the inclusion of the odd-power terms, whose ensemble averages vanish, to the $n$EPT calculation. These contributions turn out to be important for realization-by-realization comparisons \citep{Angulo2016, Pontzen2016, villaescusa-navarro2018fixpair}. Even though the $n$EPT bispectra and the SPT bispectra seem to model the $N$-body results to a similar $k_{\rm max}$ at $z\ge3$, we argue that $n$EPT modeling of spherical bispectrum is more robust than one-loop SPT because, on top of the nicer convergence, $n$EPT residual is much smoother.

However, all medications can have side effects. Both $n$EPT and SPT modeling approaches have common limitations which are most apparent for lower redshifts as demonstrated in the left panels of \reffig{WMAP}.

First, increasing order in both SPT and $n$EPT calculation does not always enhance the accuracy at lower redshift ($z \lesssim 0.5$). For example, at $z=0.5$, the accuracy of 5EPT stays below $1\%$ up to $k = 0.11\,h$/Mpc, but 6EPT starts to deviate more than $1\%$ from $k = 0.01\,h$/Mpc. One can observe similar trends for the one-loop and two-loop bispectra at $z \lesssim 0.5$. 

The second limitation is that the $n$EPT and SPT way of modeling bispectrum are sensitive to the UV (high-$k$) cutoff in the linear density field. The UV cutoff $k_{\rm cut}^{\rm UV} = (2\pi N_{\rm grid})/[(n+1)L]$ corresponds to the resolution of the GridSPT calculation. For the baseline GridSPT calculation, we use $k_{\rm cut}^{\rm UV}=1.61 \,h$/Mpc, and check the UV-cutoff dependence by repeating the calculation with $k_{\rm cut1}^{\rm UV}=1.26 \,h$/Mpc and $k_{\rm cut2}^{\rm UV}=2.14 \,h$/Mpc. The shaded region around each line in \reffig{WMAP} encloses the variation of the result when changing the $k_{\rm cut}^{\rm UV}$. Increasing UV cutoff decreases the modeled bispectrum, as shown by the hatched area in \reffig{WMAP}. The UV variation of bispectrum residual increases as the modeling involves higher-order density perturbation fields or as the nonlinearities are stronger (i.e. for the lower redshifts). At $ z \lesssim 0.5$ and for both 6EPT and two-loop SPT, the UV variation could be as large as one percent, indicated by the yellow band.

These pathological behaviors of GridSPT are shown at lower redshifts and for higher-order density fields, both of which signify when nonlinearities are large. Incorporating the back-reaction of small-scale effect on large-scale clustering, as is done in the renormalized PT and EFTofLSS \citep{Crocce2006RPT,Matsubara2008,Crocce2006memory,Crocce2012,Lazanu2016,Taruya2012,Baumann2012, Carrasco2012,Hertzberg2014}, could be a remedy, and, if that is true, a full cure calls for a field-level implementation of the renormalized PT and EFTofLSS.

\subsection{Spherical Bispectrum in General $w$CDM Cosmologies}
\label{sec:bksphDQ}
\begin{figure}[h!t]
    \centering
    \includegraphics[width=\columnwidth]{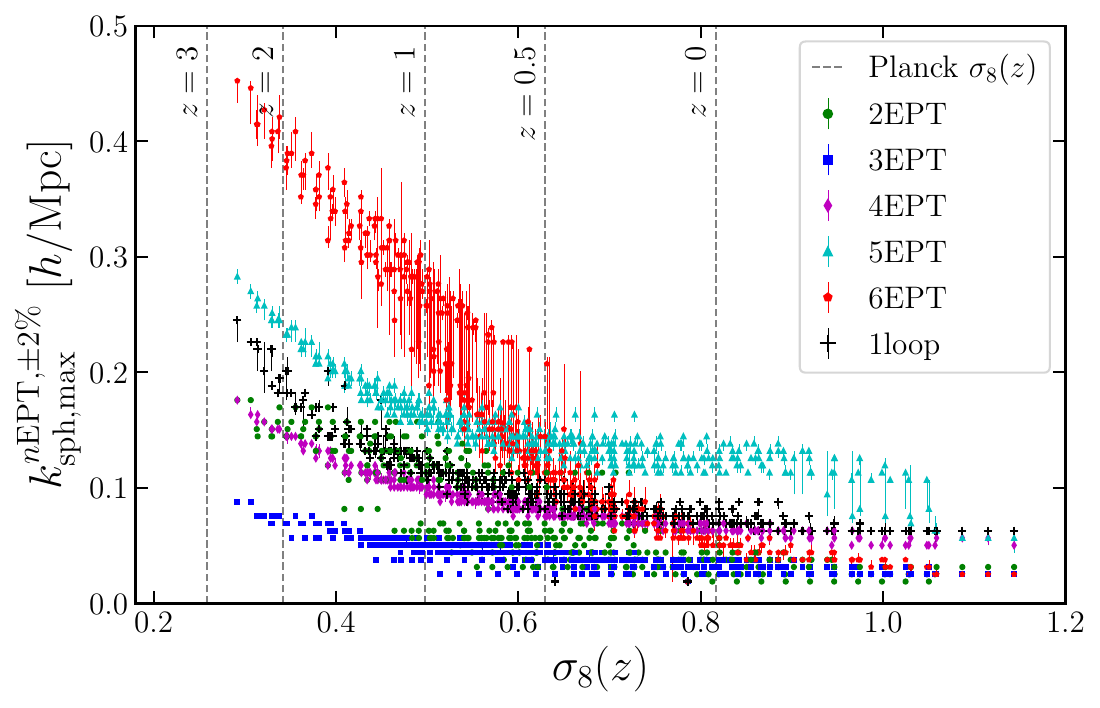}
    \caption{The anti-correlation between the maximum wavenumber, $k_{\rm sph,max}^{n{\rm EPT},\pm 2\%}$, where $n$EPT matches N-body result more than 2\% accuracy, and $\sigma_8(z)$ for Dark-Quest simulation's all 20 cosmologies at 21 redshifts $z = 0$ to $z = 1.48$. The error bars show the range of $k_{\max}$ with varying UV cutoff between $(1.26, 2.14)\, h/\mathrm{Mpc}$. The dashed lines indicate the value of $\sigma_8(z)$ in \textit{Planck} cosmology at redshifts $z = 0$, $0.5$, $1$, $1.5$, $2$, $3$.}
    \label{fig:kmaxerr2}
\end{figure}
\begin{figure*}[t]
    \centering
    \includegraphics[width=2\columnwidth]{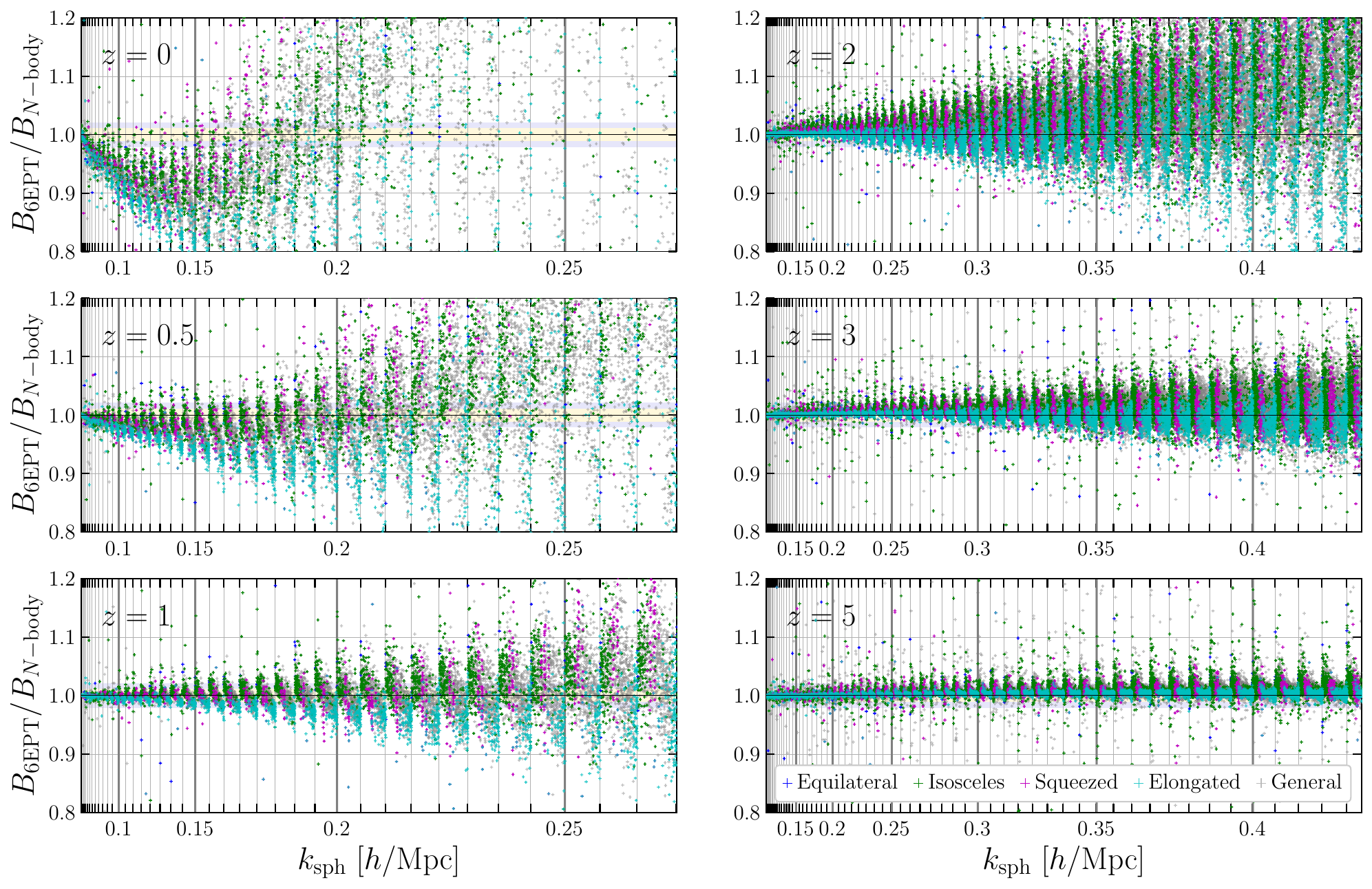}
    \caption{The ratio between 6EPT bispectrum and baseline WMAP simulation at six redshifts $z = 0, 0.5, 1, 2, 3, 5$. The 6EPT bispectra shown here are calculated with fiducial UV cutoff. The bispectra dataset is the same as that in \reffig{WMAP} but visualized in flattened index. Each chunk collects triangles with similar spherical wavenumber $k_{\rm sph}$. The interval of vertical lines equals to the bin width $k_{F}$. We distinguish configurations by different colored markers, including equilateral (blue), isosceles and o-isosceles (green), squeezed (magenta), elongated (cyan), and general triangles (grey). The yellow and lavender bands in each panel highlight the one and two-percentage error ranges.} 
    \label{fig:WMAPflat}
\end{figure*}

Can the SPT and $n$EPT modeling extend beyond the $\Lambda$CDM cosmology? We extend the comparison in the previous section with a suite of 20 Dark-Quest simulations run with general $w$CDM cosmologies. Analyzing the spherical bispectrum at 21 redshifts between $z=0$ and $z=1.48$ of all simulations confirms the conclusion we draw in the previous section also holds for these cosmologies. Of course, the limitations for $n$EPT and SPT also persist across a broad range of $w$CDM cosmologies.

Most notably, we find the strong anti-correlation between $k_{\rm sph, max}^{n{\rm EPT}, \pm 2\%}$, the maximum spherical wavenumber below which $n$EPT works better than two-percent accuracy, and the $\sigma_8(z) = \sigma_8 (z=0) D(z)$ value at each redshift snapshot, as shown by colored markers in \reffig{kmaxerr2}. This anti-correlation aligns well with the expectation that the accuracy of perturbation theory improves when the density perturbation's nonlinearity is weaker. Conversely, the stronger nonlinearities signal a break-down of the perturbative treatment: for example, at $\sigma_8(z) > 0.6$, $k_{\rm sph, max}^{6{\rm EPT}, \pm 2\%}$ goes below $k_{\rm sph, max}^{5{\rm EPT}, \pm 2\%}$  and $k_{\rm sph, max}^{4{\rm EPT}, \pm 2\%}$, which is far from the expectation of well-converging perturbation theories.

Similar to the shaded region in \reffig{WMAP}, we indicate the range of range of $k_{\rm sph, max}^{n{\rm EPT}, \pm 2\%}$ from the GridSPT calculation using the range of $k_{\rm cut}^{\rm UV}$ as the vertical errorbar. In general, the range gets much broader for higher $\sigma_8(z)$, which is consistent with the findings in the previous section. 

For comparison, we also show the maximum spherical wavenumber for one-loop SPT bispectrum as markers of black plus. We find that the one-loop SPT spherical bispectrum works accurately up to 0.1 and 0.2 $h$/Mpc at redshifts $z = 1$ and $z = 2$ in Planck cosmology. Our finding is consistent with the ensemble-mean-based comparison of spherical bispectrum done in \citep{Tomlinson2023} (see their Figure 7). We also find that 6EPT outperforms one-loop bispectrum at $\sigma_8(z) < 0.62$, which is equivalent to $z > 0.5$ in Planck cosmology. Furthermore, 6EPT spherical bispectrum could reach twice the range of validity of one-loop SPT modeling at $\sigma_8(z) < 0.5$, which corresponds to $z > 1$ in Planck cosmology.

\subsection{Modeling Bispectrum at Each Configuration}
\label{sec:bkflat}
\begin{figure*}
    \centering
    \includegraphics[width=2\columnwidth]{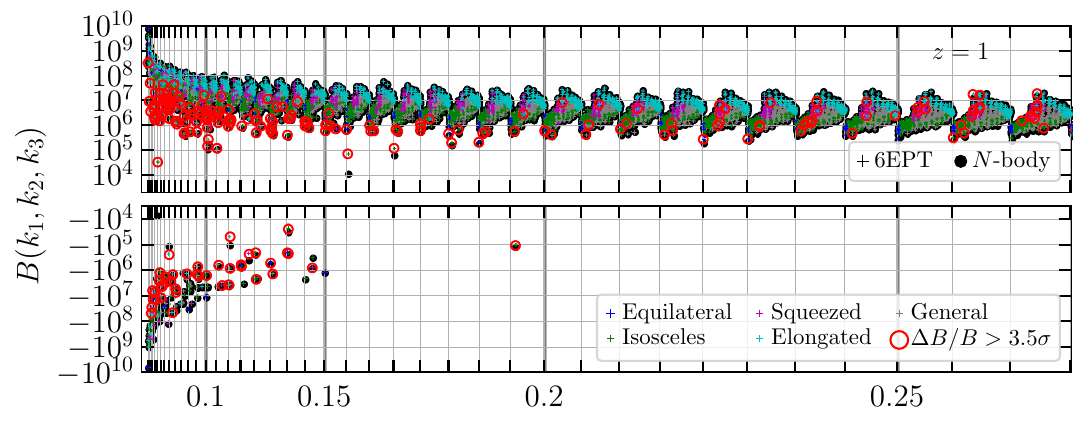}
    \includegraphics[width=2\columnwidth]{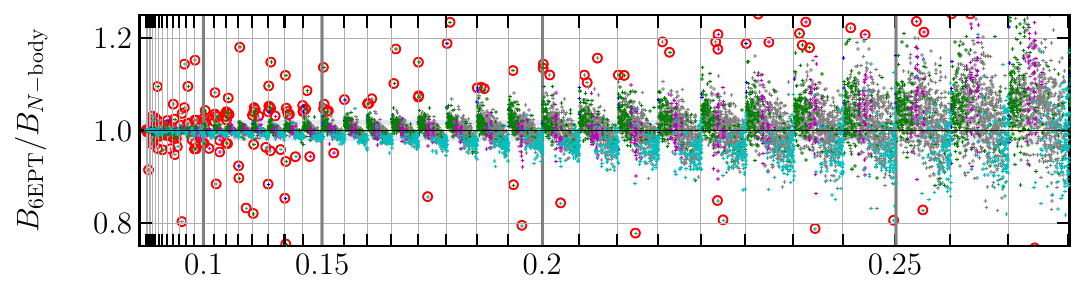}
    \includegraphics[width=2\columnwidth]{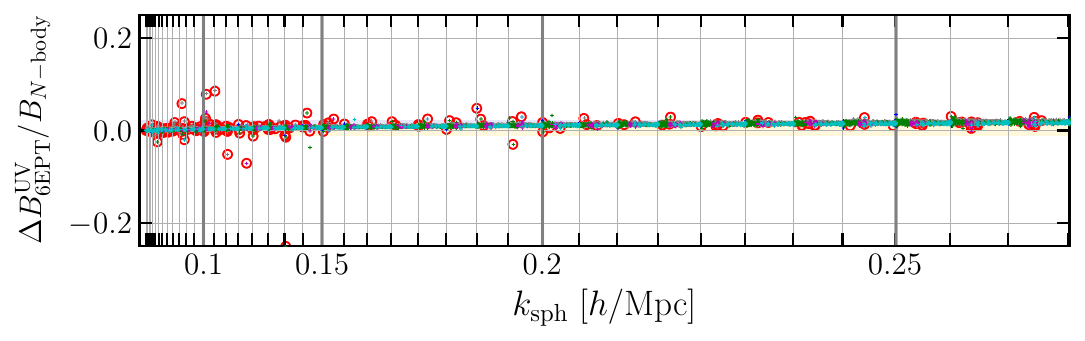}
    \caption{Comparison between the 6EPT bispectrum (with fiducial UV cutoff $k_{\rm UV,fid}$) and $N$-body results in the baseline WMAP simulation at $z = 1$. The bispectra are visualized in a flattened index, with each chunk representing a bin of fixed $k_{\rm sph}$ and bin size $k_F$. Configurations are highlighted with different colors: equilateral (blue), isosceles (green), squeezed (purple), elongated (cyan), and general (gray). The top panel shows the bispectra from the $N$-body simulation (black dots) and 6EPT (pluses). These results are measured from a single realization, so the bispectra can be negative due to cosmic variance. The middle panel displays the ratio between the 6EPT and $N$-body bispectra. The bottom panel shows the variation of the 6EPT ratio due to UV cutoffs between 1.26 and 2.14 $h/\text{Mpc}$. The yellow and lavender bands in the middle and bottom panels highlight the one and two-percentage error ranges. Across all panels, red circles highlight data points where the ratio exceeds the mean value in its chunk by more than $3.5\sigma$ in the third panel.}
    \label{fig:bkwmapflatz1var}
\end{figure*}
In the previous sections, we have examined the accuracy of SPT and $n$EPT bispectrum models in the spherical-bispectrum projection and found that both predictions provide accurate modeling for spherical bispectrum of mildly nonlinear density fields. Does this accuracy also hold for each bispectrum configuration $B(k_1,k_2,k_3)$?

To scrutinize the accuracy of the $n$EPT bispectrum of individual configuration, \reffig{WMAPflat} shows the ratio between the $n$EPT bispectrum and $N$-body result using a flattened index defined in \refsec{bkflatksph}. We stress again that the flattened index we adopt here differs from the usual one, for example used in Ref.~\citep{Scoccimarro2000}: we re-ordered the flattened index so that each chunk contains configurations within the same spherical wavenumbers $k_{\rm sph}$ bin. To examine the configuration dependence of the model accuracy while avoiding clutter, we only present the WMAP-run case and the 6EPT model with the fiducial UV cutoff. In that case, a single data point in the spherical bispectrum in \reffig{WMAP} represents a weighted average of all the data points within the corresponding {\it chunk} in \reffig{WMAPflat}, where the weights are determined by the number of Fourier triangles contributed to each point.

While \reffig{WMAP} shows that 6EPT models the spherical bispectrum to a sub-percent accuracy on quasi-linear scales, the typical error of the bispectrum of each configuration shown in \reffig{WMAPflat} is larger by a factor of a few! For example, on large scales where the 6EPT can model the spherical bispectrum to a sub-percent in \reffig{WMAP}, errors shown in \reffig{WMAPflat} for individual configurations fluctuate a lot and reach nearly ten percent. 
The residual errors also show some tendency depending on its triangular configurations. 6EPT tends to over-predict the bispectra of equilateral and isosceles triangles while under-predicting the bispectra of elongated triangles. Such a wide scatter of errors has been canceled and has remained invisible to the spherical bispectrum. That is, when averaged over the points in the chunk, the errors of different configurations cancel out. This is the reason why the error of spherical bispectrum in \reffig{WMAP} is much smaller. 

In addition to the scatter in the residual error, we observe that for some (about $1.5\%$ of the total at $z=1$) configurations, the residual error significantly exceeds that of other points in the same $k_{\rm sph}$ bin. \reffig{bkwmapflatz1var} further elaborates on this point for the $z=1$ case. The top panel of \reffig{bkwmapflatz1var} shows the bispectra from the 6EPT (plus marker) and $N$-body simulations (black dots). First of all, note that while the theoretical prediction of the ensemble average of the PT bispectrum is positive, the bispectrum itself is not a positive definite quantity so its value in a realization can be negative. Indeed, certain data points at lower $k_{\rm sph}$ in the top panel of \reffig{bkwmapflatz1var} are negative. 

\begin{figure}
    \centering
    \includegraphics[width=\columnwidth]{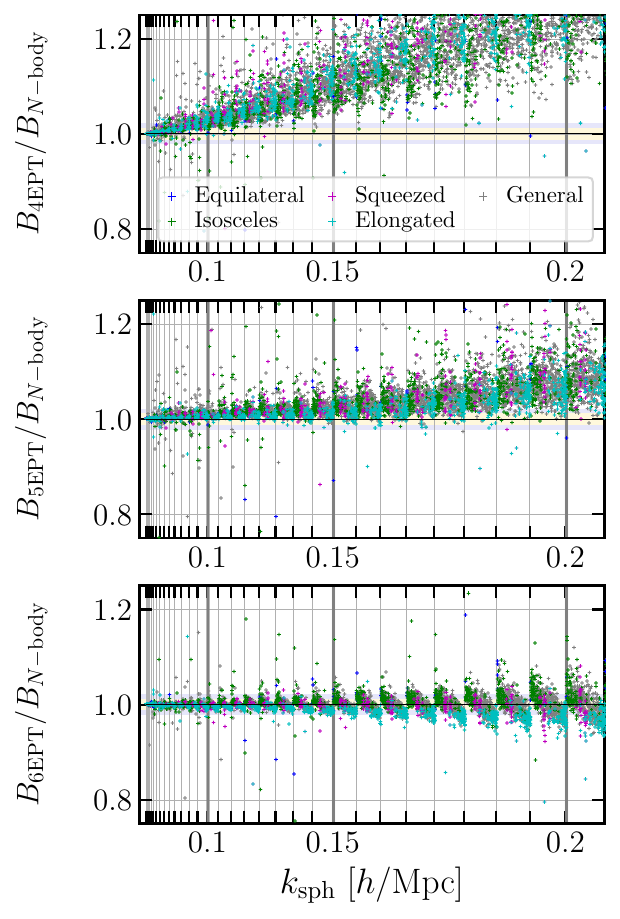}
    \caption{Comparison between 4EPT (the top panel), 5EPT (the middle panel), and 6EPT bispectrum (the bottom panel) with fiducial UV cutoff $k_{\rm UV,fid}$ and $N$-body results in the baseline WMAP simulation at $z = 1$. The legends are the same as that in \reffig{bkwmapflatz1var}.}
    \label{fig:bkwmapflatz1var456}
\end{figure}
\begin{figure*}
    \centering
    \includegraphics[width=2\columnwidth]{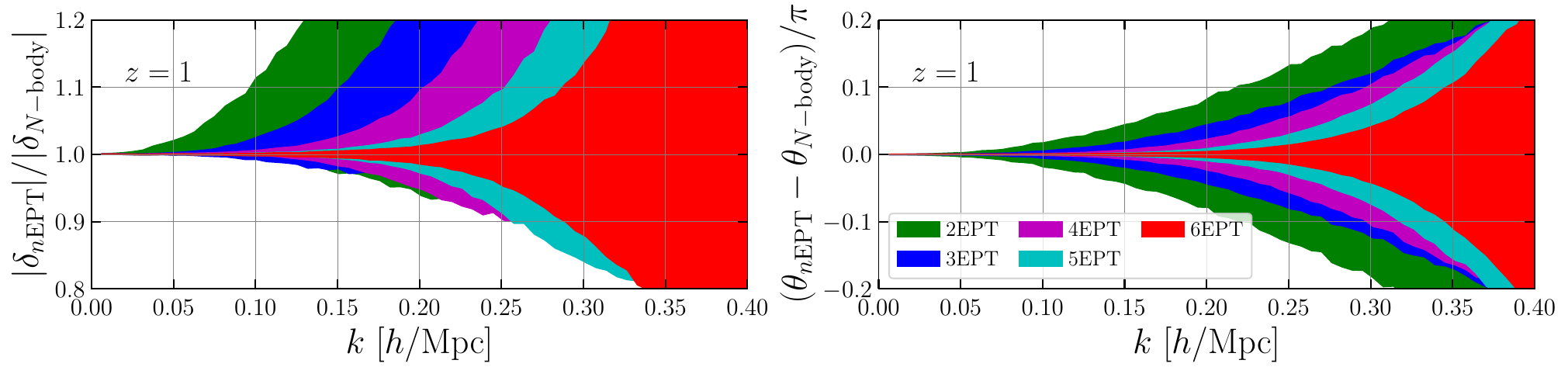}
    \includegraphics[width=2\columnwidth]{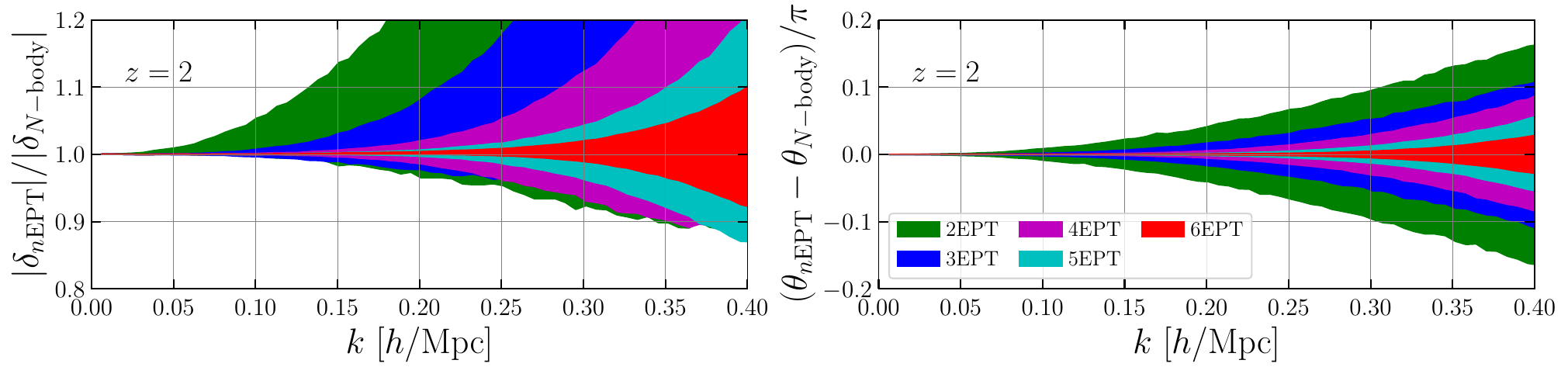}
    \includegraphics[width=2\columnwidth]{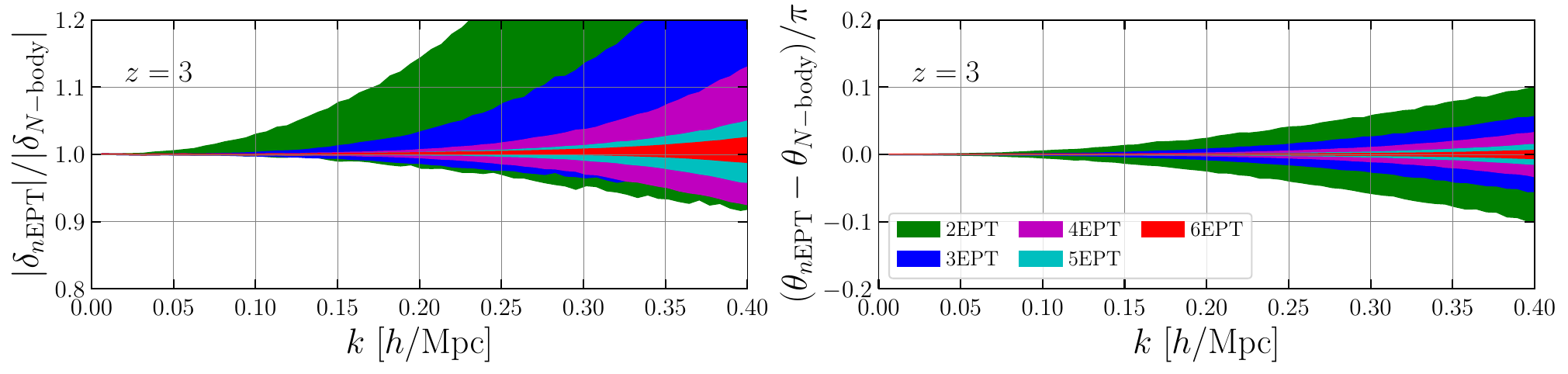}
    \caption{The comparison between the density fields in N-body simulations and that forward modeled by $n$EPT (GridSPT) at $z = 1, 2, 3$. 
    {\it Left panel:} The ratio of the modulus of each Fourier density mode. {\it Right panel:} The phase difference of each Fourier density mode. The colored regions represent the 60\% quantile of the data, which is computed among all the Fourier modes within a $k$-bin, where the bin size is $k_F$.}
    \label{fig:fieldz1}
\end{figure*}

The ratios between 6EPT and $N$-body bispectra, shown in the middle panel of \reffig{bkwmapflatz1var}, mirror those in the bottom left ($z=1$) panel of \reffig{WMAPflat}. In that panel, we also highlight the outliers at each $k_{\rm sph}$-bin chunk by red. We define outliers as the data points deviating from the chunk's mean by more than $3.5\sigma$, with $\sigma$ representing the normalized mean absolute deviation (NMAD) within each $k_{\rm sph}$ chunk. Note that the distribution of residual is highly non-Gaussian so that $3.5\sigma$ outliers at each side consists of $1.5\%$ of the total configurations.
Comparing the circled data points in \reffig{bkwmapflatz1var} on large scales, we find a pattern among these outliers that their bispectra, serving as the denominator in fractional error calculation, are significantly lower—by one or two orders of magnitude—compared to other data points in the chunk. We also find that the outliers in the DarkQuest simulation with similar $\sigma_8(z)$ appears not for the same triangular configurations shown in \reffig{bkwmapflatz1var}. These findings suggest that the outliers in the fractional error may not be as troublesome as they seem, and we surmise that by averaging over a sufficient number of realizations would remove the outliers.

Finally, the bottom panel of \reffig{bkwmapflatz1var} shows that the variation of bispectrum of each configuration due to UV cutoff between (1.26, 2.14) $h$/Mpc is less than or around one percent among all the scales we presented here. So the UV variation becomes a subdominant factor in the systematic errors of 6EPT bispectrum for each configuration.

\reffig{bkwmapflatz1var456} shows the ratio between the 4EPT (the top panel), 5EPT (the middle panel), and 6EPT bispectra (the bottom panel) of each configuration against $N$-body results at $z=1$. The mean ratios for each chunk of 4EPT and 5EPT deviate from one around $k = 0.1\,h$/Mpc and $k=0.15\,h$/Mpc, respectively, which is consistent with the bottom left panel of \reffig{WMAP}. Another notable feature is the decrease in scattering within each chunk as the order of $n$EPT increases, indicating that higher-order $n$EPT indeed improves the modeling of the bispectrum for individual configurations. Such information is absent from the spherical bispectrum in \reffig{WMAP}.

\subsection{The Residual of the $n$EPT Modeling of Fourier-Space Density Field}
Why is the matter bispectrum modeling residual for each configuration several times larger than the spherical bispectrum at a fixed scale? 
Since the bispectrum depends on both the moduli and phases of the three density modes, the residual of $n$EPT bispectrum is determined by the residual of the density fields at the field level.
To take a cursory look, we compare directly the nonlinear density contrast $\delta=|\delta|e^{i\theta}$ of $N$-body simulation with the $n$EPT prediction. 

\reffig{fieldz1} shows the 60\% quantile range of the density-moduli ratio (the left panels) and phase difference (right panels) between $\delta_{n\rm EPT}(\vk)$ and $\delta_{N-{\rm body}}(\vk)$. The plot is for the baseline WMAP run, and we compute the 60\% spread within a Fourier bin of size $\Delta k=k_F=0.00628~h/$Mpc.
Again, $\delta_{n\rm EPT}(\vk,z)=\sum_{i=k}^n [D(z)]^k\delta^{(k)}(\vk)$ is the $n$EPT density contrast computed from GridSPT. \reffig{fieldz1} shows the results from 2EPT to 6EPT (colors as indicated in the legend) at $z = 1, 2, 3$ (from top to bottom). We emphasize that the result is computed based on the distribution of the ratios in each bin, which is different from the binned average. 

Since the power spectrum is the square of the density modulus, the density-moduli ratio plot (left panel of \reffig{fieldz1}) essentially shows the spread of the power spectrum in a mode-by-mode manner. Surprisingly, we find that the $n$EPT density moduli have residuals larger than the averaged power spectrum by a factor of few, even on large scales where we have found a sub-percent accuracy of $n$EPT power spectrum in \cite{Wang2023nEPT}. For example, the residual of 5EPT power spectrum stays within the sub-percent level to $k=0.25\,h$/Mpc at $z = 1$ (see Fig. 1 of \citep{Wang2023nEPT}), but the 60\% quantile of the moduli difference could be $\pm 10\%$. Therefore, this plot also indicates that the accuracy of averaged summary statistics does not guarantee the mode-by-mode level match.

The phase residuals (right panel of \reffig{fieldz1}) for all wavenumbers have a smaller scatter than the moduli residual, but the match is still worse than what has been reported using summary statistics. For example, the cross-correlation coefficient used in Ref.~\citep{Taruya2018Grid} is defined as 
\be
R_{\rm corr}^{(n)}(k) = \frac{\sum{\rm Re}[\delta_{N{\rm-body}}(\vk)\delta_{n{\rm EPT}}(-\vk)]}{\(\sum|\delta_{N{\rm-body}}(\vk)|^2\)^{1/2}\(\sum|\delta_{n{\rm EPT}}(\vk)|^2\)^{1/2}}\,,
\ee
where all sums add the Fourier modes within the bin. Fig. 10 of \citep{Taruya2018Grid} shows that for $k \le 0.2\,h$/Mpc at $z = 1$, the cross-correlation coefficient between both 4EPT and 5EPT against $N$-body results stay very close to unity without much scatter. The denominator is the bin-averaged power spectra, which are rather smooth, so the lack of scatter in the cross-correlation coefficient reflects that the numerator, for which scatter from both moduli and phase can contribute, is also smooth when taking an average within the bin. Again, the fact that cross-correlation coefficient at a given $k$ close to one does not guarantee the small phase residual at each Fourier mode. Indeed, we find that the largest phase residual of 5EPT could reach $0.4\pi$ at $k=0.2 h$/Mpc at $z = 1$. While the Fig. 10 of \citep{Taruya2018Grid} shows that the 4EPT and 5EPT perform almost identically for the cross-correlation coefficient, from the top right panel of \reffig{fieldz1}, we find that the 5EPT clearly outperforms 4EPT in the modeling of phase on $k<0.2~h/$Mpc at $z=1$.

For both modulus residual and phase residual plots, the large scatter in the field-level residuals can be mitigated through binning, which reduces the residuals in the binned power spectrum, spherical bispectrum, or the cross-correlation coefficient. As the binning includes more number of modes, the field-level residuals we show in \reffig{fieldz1} are more likely to average out to yield smaller residuals of summary statistics. That is, a coarse-grained binning strategy such as spherical bispectrum could better hide the field-level residuals compared to the bispectrum for a fixed configuration. This is the reason behind the larger scatter in \reffig{WMAPflat} than \reffig{WMAP}. As the bottom panel of \reffig{bkwmapflatz1var} shows, changing the UV curoff only gives a minor correction to the scatter.

For the statistical field-level analysis using the GridSPT, the distribution of residuals that we show in \reffig{fieldz1} must be quantified to contstruct the field-level likelihood function. We will present the result elsewhere in the future publication.

\section{Discussion}
\label{sec:conclusion}
In this paper, we compare the $n$EPT modeling of the nonlinear matter bispectrum in real space to the $N$-body simulation results to confirm the success of $n$EPT extends beyond modeling the nonlinear matter power spectrum as shown in Ref.~\cite{Wang2023nEPT}. 

Starting from the GridSPT outcome, $n$EPT first sum up the density contrast up to the $n$-th order and then compute the summary statistics such as the matter bispectrum. This is an alternative to the conventional loop-by-loop calculation of the SPT bispectrum. We show that $n$EPT bispectrum has following advantages over the SPT bispectrum: (A) Better regulated convergence when successively increasing the order $n$, (B) Much smoother residual, and (C) Substantial increase of the quasi-linear scales.

Although 5EPT and 6EPT bispectrum model significantly extend the quasi-linear scale, they are most sensitive to the UV cutoff in the initial linear density field (or the grid size adopted to compute GridSPT density contrast). As we can see from \reffig{WMAP}, changing the UV cutoff by about a factor of two (from 1.26 $h$/Mpc to 2.14 $h$/Mpc) yields a percent-level variation of 6EPT bispectrum around $k_{\rm max}$. Since the residual curve for $z>2$ varies rather shallowly, a percent-level change could shift the $k_{\rm max}$ by a larger factor.

While we find $n$EPT models spherical bispectrum accurately more than percent-level on quasi-linear scales, the typical error of bispectrum of individual configuration is several times larger than that of the spherical bispectrum. This is because the error of bispectrum of different configurations are cancelled out when get averaged to get spherical bispectrum. We prove this by showing that the large errors observed in the bispectrum of specific configurations originate from the spread between $n$EPT density contrast and $N$-body results. These field-level residuals can be mitigated through the binning of summary statistics, such as the measured $n$EPT power spectrum reported in \citep{Wang2023nEPT}, the cross-correlation coefficient calculated through the binned auto- and cross-power spectrum \citep{Taruya2018Grid}, or the $n$EPT spherical bispectrum \citep{Tomlinson2023} used in this study. This suggests that the dynamic range of $n$EPT, and essentially GridSPT, for field-level inferences is more limited than that for $n$-point correlation functions (refer to \citep{Schmidt2021b, Nguyen2021, Kostic2023, Stadler2023, Tucci2023, Nguyen2024} for examples of using forward modeling at field-level by Lagrangian perturbation theory). Further studies are in order to quantify the statistics of the forward-modeling residuals, which can be used to set up the likelihood function for the field-level cosmological analysis.

The poor convergence we observe at low redshift ($z < 0.5$) where nonlinearities are strong, the UV-cutoff dependence of the $n$EPT outcome, and the field-level residual of the density field may all be stemmed from the limitations of standard perturbation theory, which neglect (A) vorticity, and (B) higher-order cumulants of phase space distribution beyond momentum ($n=1$). The $N$-body simulations solve full Vlasov--Poisson equation without these assumptions.
Therefore, a theory beyond SPT must include all higher-order cumulants, which are generated by shell crossing in the non-linear regions \citep{Pueblas2009}, into Vlasov--Poisson equation. Recently proposed {\it Vlasov Perturbation Theory} (VPT) \citep{Garny2023a, Garny2023b} aims to achieve this goal. Their solutions suppress the UV power in linear density modes and kernel of non-linear density modes compared to SPT solutions, which are potentially helpful to reduce the UV sensitivity and also improve the accuracy of the field-level modeling.

In order to apply the $n$EPT framework to the analysis of the galaxy power spectrum and bispectrum, we have to incorporate the perturbative galaxy bias expansion \citep{Desjacques2018bias} into GridSPT. Further details on the proper renormalization of bias operators will be addressed in a forthcoming paper \citep{Wang2024biasrenorm}. For modeling the galaxy power spectrum and bispectrum in redshift space, $n$EPT can utilize the velocity field from each grid in GridSPT. This approach includes implementing non-linear redshift-space distortion mapping \citep{Taruya2022RSD} and accounting for selection bias \citep{Desjacques2018} directly at the field level. Leveraging the efficiency of GridSPT, it is also feasible to calculate the covariance matrix of the galaxy power spectrum and bispectrum, incorporating the survey geometry from a large number of realizations \citep{Taruya2021Grid}. We anticipate that $n$EPT fundamentally changes the way of modeling summary statistics from perturbation theory in Eulerian space and enable more accurate modeling of galaxy summary statistics on smaller scales and enhance the extraction of cosmological information from galaxy surveys.

\appendix
\section{Measure the bispectrum components in GridSPT with minimum steps}
\label{app:bijkoptimal}

To measure $n$EPT and SPT bispectrum from density perturbation fields at each order in GridSPT, we need to measure all components appearing in \refeq{nEPTbk} and \refeq{SPTbk}. In this section, we introduce a novel strategy to quickly measure all these components by taking advantage of the permutational symmetry. We denote the order of $n$EPT as $n$ and the number of redshifts as $N_z$. 

For estimating the SPT and nEPT bispectrum, we first compute each component at $z=0$ and use the power of linear growth factor $D(z)$ to rescale the components to targeted redshift and then sum up all components,
\bea
\<\dn i(\vk_1)\dn j(\vk_2) \dn k(\vk_3)\> D^{i+j+k}(z).
\eea
Here, $i,j,k \le n$. The na\"ive way of computing all components up to $n$ requires us to run the estimator for $n^3$ times. We can also combine the components above to estimate the $n$EPT bispectrum.

We reduce the number of bispectrum estimation calculations by using the index-permutation symmetry as follows. Let us begin by classifying the bispectrum components into three groups and estimating the number of required calculations.
\begin{enumerate}
    \item $B_{iii}$, $n$ times
    \item $B_{iij}$ + 2 perms, $3n(n-1)$ times
    \item $B_{ijk}$ + 5 perms, $n(n-1)(n-2)$ times
\end{enumerate}
We cannot change the times of estimation in $B_{iii}$. For $B_{iij}$, we can use the following tricks. We first add $\dn i$ and $\dn j$, denoted as $\delta_{i+j}$, and then measure its bispectrum, which is
\bea
B_{i+j}&\equiv&\<\delta_{i+j}(\vk_1)\delta_{i+j}(\vk_2)\delta_{i+j}(\vk_3)\>^\prime
\nn\\
&=&
B_{iii} + B_{jjj}
\nn\\
&+& \(B_{ijj} + 2\;{\rm perms}\) + \(B_{iij} + 2\;{\rm perms}\)
\nn\\
\eea

To separate the last two terms, we need to measure the bispectrum from another linear combination of $\dn i$ and $\dn j$. The easiest way is to compute the bispectrum of $\dn i$ minus $\dn j$, denoted as $\delta_{i-j}$, which is
\bea
B_{i-j}&\equiv&\<\delta_{i-j}(\vk_1)\delta_{i-j}(\vk_2)\delta_{i-j}(\vk_3)\>^\prime
\nn\\
&=&
B_{iii} - B_{jjj}
\nn\\
&+& \(B_{ijj} + 2\;{\rm perms}\)
 - \(B_{iij} + 2\;{\rm perms}\)
\nn\\
\eea

Then we have
\bea
\(B_{ijj} + 2\;{\rm perms}\) = \frac12\[B_{i+j} + B_{i-j}\] - B_{iii}
\\
\(B_{iij} + 2\;{\rm perms}\) = \frac12\[B_{i+j} - B_{i-j}\] - B_{jjj}
\eea 

Since we know $B_{iii}$ and $B_{jjj}$, we can get $B_{ijj}$ and $B_{iij}$ along with their permutations from the two equations above. What we have done is to measure $B_{i+j}$ and $B_{i-j}$ for all possible $i > j$. In total, measuring $B_{iij}$-type bispectra along with the two permutations in this way requires running estimator only for $n(n-1)$ times, which is smaller by a factor of three compared to the naive algorithm.

Lastly, to measure $B_{ijk}$ along with its five permutations, we can first add $\dn i$, $\dn j$, and $\dn k$, denoted as $\delta_{i+j+k}$, and then measure its bispectrum, which is
\bea
&&\<\delta_{i+j+k}(\vk_1)\delta_{i+j+k}(\vk_2)\delta_{i+j+k}(\vk_3)\>^\prime
\nn\\
&=&
B_{iii} + B_{jjj} 
\nn\\
&& + \(B_{ijj} + 2\;{\rm perms}\)
   + \(B_{iij} + 2\;{\rm perms}\)
\nn\\   
&& + \(B_{ikk} + 2\;{\rm perms}\)
   + \(B_{iik} + 2\;{\rm perms}\)
\nn\\   
&& + \(B_{jkk} + 2\;{\rm perms}\)
   + \(B_{jjk} + 2\;{\rm perms}\)
\nn\\   
&& + \(B_{ijk} + 5\;{\rm perms}\)
\eea
Since we have measured all terms other than $B_{ijk}$, we can get $B_{ijk}$ along with five permutations by only running estimator once for given \{$i,j,k$\}. Spanning over all possible unpermuted \{$i,j,k$\} requires $n(n-1)(n-2)/6$ times of measurement, which is smaller by a factor of six compared to the naive algorithm.

In summary, with the new algorithm, measuring all the three groups of bispectrum components, we need to run the estimator for 
\bea
&& n + n(n-1) + n(n-1)(n-2)/6 
\nn\\
&& = n(n+1)(n+2)/6
\eea
times.

Throughout this paper, we use $n=6$ for assessing the 6EPT and 2-loop bispectrum derived from density fluctuations within GridSPT. The conventional algorithm necessitates $n^3 = 216$ times of measurement. However, with the introduction of our advanced algorithm, this requirement diminishes to 56, marking a reduction by approximately a factor of four.

\section*{Acknowledgement}
DJ and ZW was supported at Pennsylvania State University by NSF grant (AST-2307026). DJ is supported by KIAS Individual Grant PG088301 at Korea Institute for Advanced Study.
TN is supported by MEXT/JSPS KAKENHI Grant Number JP20H05861, JP21H01081, JP22K03634, JP24H00215 and JP24H00221.
KO is supported by JSPS KAKENHI Grant Number JP22K14036 and JP24H00215.
This research has made use of NASA's Astrophysics Data System and adstex (\url{https://github.com/yymao/adstex}). ZW wants to acknowledge Joseph Tomlinson for sharing his code of efficient paralyzed bispectrum estimator in JULIA language. ZW wants to acknowledge Fan Zou for the helpful discussion on NMAD used in \reffig{bkwmapflatz1var}. ZW also wants to acknowledge Thomas Fl${\rm \ddot o}$ss for useful discussion on the comparison between Eulerian and Lagrangian perturbation theory. The authors want to acknowledge Yukawa Institute for Theoretical Physics, Kyoto University for hosting the workshop ``Revisiting cosmological non-linearities in the era of precision surveys" (YITP CNLWS 2023: YITP-T-23-03) which facilitates the completion of this work.

\bibliography{main}

\end{document}